\documentclass[12pt]{amsart} 
\input xy \xyoption{all} \xyoption{poly} 
 
\input{epsf}             
\def\epsfcenter#1{{\vcenter{\hbox{\epsfbox{#1}}}}} 

\newcommand{\SL}{{\rm SL}} 
 
\newcommand{\SU}{{\rm SU}}

\newcommand{\SO}{{\rm SO}}

\newcommand{\dd}{{\rm d}} 

\newcommand{\C}{{\mathbb C}}  
  
\newcommand{\nfactor}{{\frac1{2\pi^2}}}

\newtheorem{defn}{Definition}  
   
\newtheorem{lemma}{Lemma}  
  
\newtheorem{theorem}{Theorem}

\renewcommand{\H} {{H}}  
  
        \newcommand{\be}{\begin{equation}}  
        \newcommand{\ee}{\end{equation}}  
        \newcommand{\ba}{\begin{eqnarray}}  
        \newcommand{\ea}{\end{eqnarray}}  
        \newcommand{\ban}{\begin{eqnarray*}}  
        \newcommand{\ean}{\end{eqnarray*}}  
        \newcommand{\barr}{\begin{array}}   
        \newcommand{\earr}{\end{array}}

\newcommand{\FourX}[4]  
{  
\xymatrix{\ar@{-}[dr]^{#1} & & \ar@{-}[dl]^{#2} \\  
 & *{\bullet} & \\  
\ar@{-}[ur]^{#3} & & \ar@{-}[ul]^{#4}\\}  
}  
  
\newcommand{\DoubleY}[5] 
{  
\xymatrix{\ar@{-}[dr]^{#1} & & & \ar@{-}[dl]_{#2} \\  
 & *{\bullet} \ar@{-}[r]^{#5}& *{\bullet} &\\  
\ar@{-}[ur]_{#3} & & & \ar@{-}[ul]^{#4}\\}  
}  
  
\newcommand{\monogon}[1]  
{  
\xymatrix{ *{\bullet}  
\ar@  
{-}  
@(ul,dl)  
[]  
_{#1}  
\\}  
}  
  
\newcommand{\bigon}[2]  
{  
\xymatrix{ *{\bullet} 
\ar@{-} 
@/^1pc/ 
[r] 
^{#1} 
\ar@{-} 
@/_1pc/ 
[r] 
_{#2} 
& 
*{\bullet} 
\\} 
} 
 
\newcommand{\thetagraph}[3]  
{ 
\xymatrix{ *{\bullet} 
\ar@{-} 
@/^1.5pc/ 
[r] 
^{#1} 
\ar@{-} 
@/_1.5pc/ 
[r] 
_{#2} 
\ar@{-} 
[r]^{#3} 
& 
*{\bullet} 
\\} 
} 
 
\newcommand{\unigon}[1]  
{ 
\xymatrix{ *{\bullet} 
\ar@{-} 
[r]^{#1}  
& 
*{\bullet} 
\\} 
}

\newcommand{\fourtheta}[4]  
{ 
\xymatrix{ *{\bullet} 
\ar@{-} 
@/^1.5pc/ 
[r] 
^{#1} 
\ar@{-} 
@/_1.5pc/ 
[r] 
_{#2} 
\ar@{-} 
@/^/ 
[r]^{#3} 
\ar@{-} 
@/_/ 
[r]^{#4} 
& 
*{\bullet} 
\\} 
}

\begin{document} 
 
\title{Integrability for Relativistic Spin Networks}   
\author{John C.\ Baez} 
\address{Department of Mathematics, University of California, 
Riverside CA 92507, USA} 
\author{John W.\ Barrett } 
\address{School of Mathematical Sciences, University 
Park, Nottingham NG7 2RD, UK}  
  
 
\begin{abstract}   
The evaluation of relativistic spin networks plays a fundamental role in
the Barrett-Crane state sum model of Lorentz-\break ian quantum gravity
in 4 dimensions.  A relativistic spin network is a graph labelled by
unitary irreducible representations of the Lorentz group appearing in
the direct integral decomposition of the space of $L^2$ functions on
three-dimensional hyperbolic space.  To `evaluate' such a spin network
we must do an integral; if this integral converges we say the spin
network is `integrable'.  Here we show that a large class of
relativistic spin networks are integrable, including any whose
underlying graph is the 4-simplex (the complete graph on 5 vertices).
This proves a conjecture of Barrett and Crane, whose validity is
required for the convergence of their state sum model.
\end{abstract}
 
\maketitle  
  
\section{Introduction} 
 
In formulating a state sum model for 4-dimensional Lorentzian quantum 
gravity, Barrett and Crane \cite{BC2} used the notion of a  
`relativistic spin network', which is simply a graph with edges labelled 
by nonnegative real numbers.  These numbers parametrize a certain 
class of irreducible unitary representations of the Lorentz group. 
The model involves a triangulation of spacetime, and for each 4-simplex  
one must calculate an amplitude associated to a relativistic spin network  
whose underlying graph is the complete graph on 5 vertices: 
\medskip 
$$\xy/r3pc/: 
{\xypolygon5~*{\bullet}}, 
"1";"3"**@{-}, 
"2";"4"**@{-}, 
"3";"5"**@{-}, 
"4";"1"**@{-}, 
"5";"2"**@{-} 
\endxy 
$$ 
\medskip 
 
To calculate this amplitude one must do an integral.  In \cite{BC2} this 
integral was conjectured, but not proven, to converge.  Here we prove  
that it does in fact converge.   More generally, one can associate a 
similar integral to any relativistic spin network.  To `evaluate' a 
relativistic spin network one must do this integral.   We prove that this 
integral converges whenever the underlying graph of the spin network lies  
in a certain large class.  This class of graphs includes the tetrahedron and 
all graphs obtained from it by repeatedly adding an extra vertex 
connected by at least 3 edges to the existing graph.  Thus in particular 
this class includes the complete graph on 5 vertices.   
 
The results of the paper are formulated in Section \ref{eval} and proved 
in Section \ref{proofs}. These sections are developed in a 
self-contained manner, so that the reader interested in the mathematical 
details can simply start there. 
 
The remainder of the introduction contains some background material 
placing the results in their mathematical and physical contexts. The 
mathematical part of the introduction explains how the integrals 
we consider arise from the representation theory of the Lorentz group.   
The physics part of the introduction sketches how this representation  
theory has been used in constructing models of fundamental physics. 
 
\subsection{Mathematical context} 
 
Calculations in the representation theory of compact Lie groups are 
conveniently expressed in terms of diagrams. Let $a$, $b$, ..., $n$ and 
$p$, $q$ ,... , $z$ be representations of the group and $A\colon a\otimes 
b\otimes \ldots\otimes n\to p\otimes\ldots\otimes z$ an intertwining 
operator (a linear map which commutes with the action of the group). This  
can be represented by a diagram:
   
\medskip
$$\epsfysize=1.5in \epsfbox{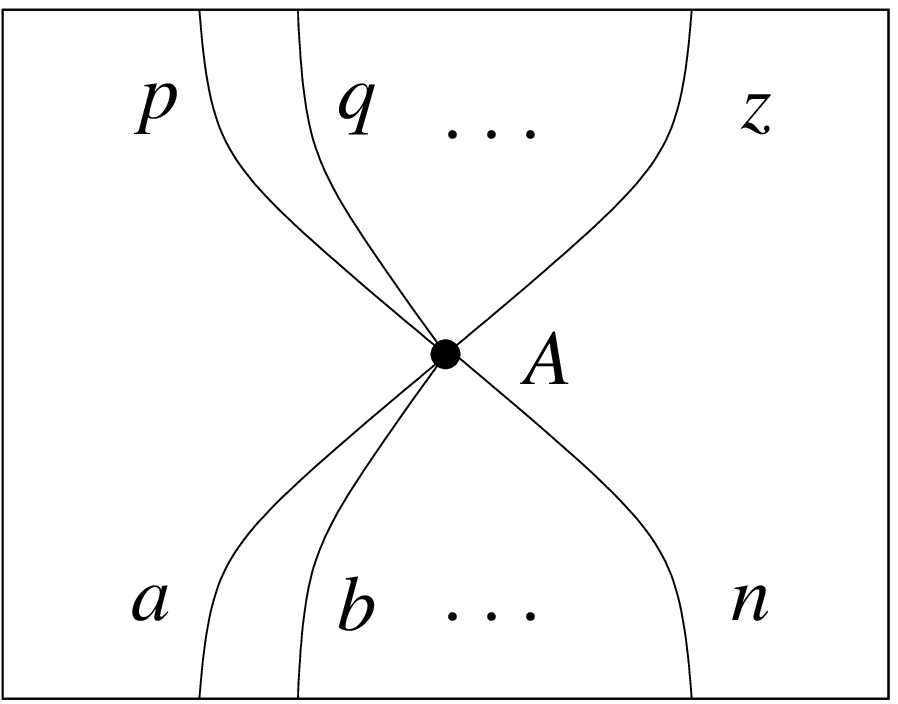}$$ 
\medskip
 
\noindent
Usually we fix some particular operators and represent them by diagrams 
like this; these operators are called the \textit{elementary operators}, or 
sometimes \textit{vertices}, corresponding to the fact that the graph in 
the diagram has just one vertex. 
  
The identity operator is represented by a vertical line with no vertex. 
Tensoring corresponds to joining diagrams together horizontally, and 
composition of operators corresponds to joining diagrams vertically. A 
trace on a representation, say $a=p$ in the above diagram, is 
represented by joining the edge for $a$ round in a loop to the edge 
for $p$.  These processes allow us to build various operators from the 
elementary ones, with the description in terms of elementary 
operators being captured by a diagram: a graph with vertices 
labelled by elementary operators and a certain number of free 
ends on the bottom and top, corresponding to `inputs' and `outputs':
 
\medskip
$$\epsfysize=1.9in\epsfbox{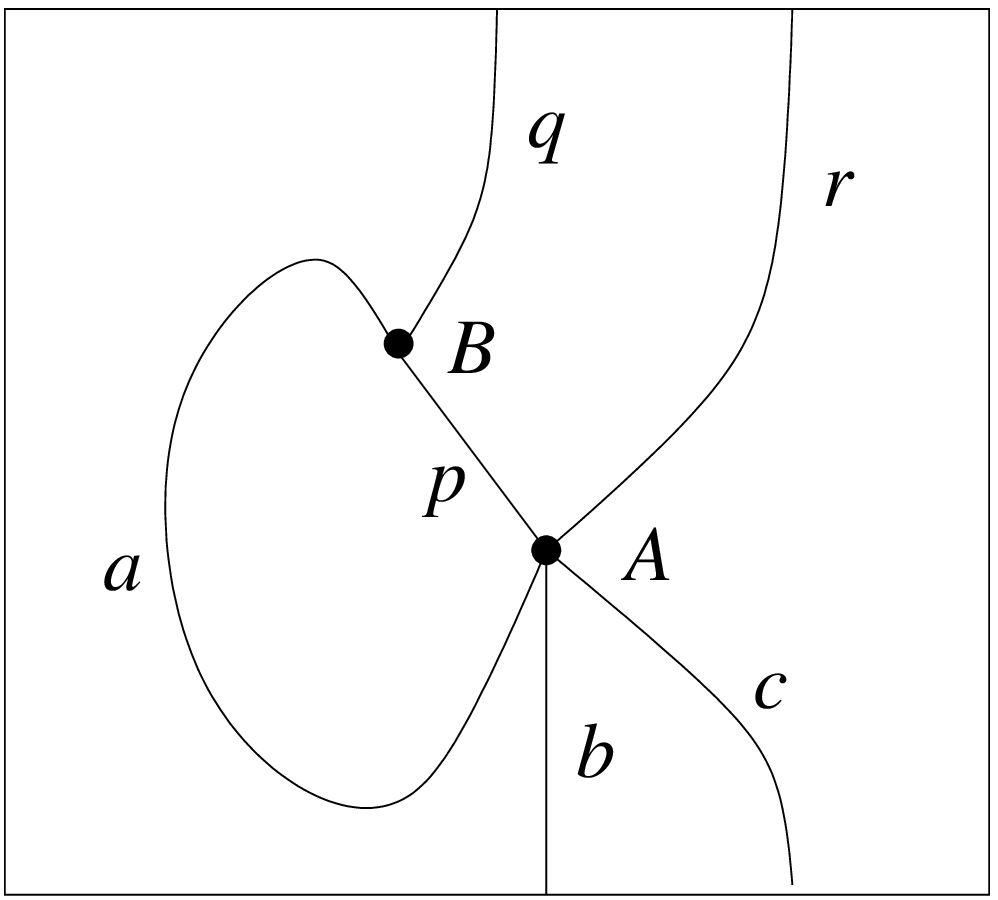}$$
\medskip

The power of this method comes from the fact that relations between 
these intertwining operators correspond, in many cases, to deformations 
of the diagram which can be interpreted as moving the vertices and edges 
in either two- or three-dimensional space (isotopies).  This theory was 
developed to its fullest extent in the generalisation from compact Lie 
groups to a certain class of Hopf algebras, particularly the quantum 
groups.  Since a diagram with no free ends corresponds to an 
intertwining operator from the trivial representation to itself, i.e.\ 
essentially just a complex number, such Hopf algebras yield invariants 
of knots and graphs embedded in three-dimensional space. 
 
However in this paper we are concerned with a generalisation in a 
different direction, namely from compact to non-compact Lie groups, and 
study a particular class of unitary representations of the Lorentz 
group, $\SO_0(3,1)$.  The subscript here indicates the connected 
component of $\SO(3,1)$ that contains the identity.  This is the group 
covered by $\SL(2,\C)$. 
 
We restrict attention to a particularly simple class of elementary operators 
(vertices) which are invariant under all permutations of the 
edges in the diagram. In essence, as will be explained below, the 
composite intertwining operators depend only on the underlying graph of the 
diagram and not on the way in which it is drawn on the plane; the 
knot-theoretic considerations are absent. The difficulties in the theory 
are of a different nature. The trace cannot be defined in all 
circumstances since the representations are 
infinite-dimensional, and the ordinary trace of the identity operator is 
infinite. However we will show that there is another perfectly good notion of 
trace provided that the diagram is sufficiently connected. 
 
The representations of the Lorentz group considered here are as 
follows. There is one for each real number $p\ge0$. 
The Hilbert space of this representation (also denoted $p$) is 
the space of solutions to the equation $\nabla^2 f= -(p^2+1)f$ on 
three-dimensional hyperbolic space, $\H$, which have square-integrable 
boundary data on the sphere at infinity \cite{BC2}.  
This Hilbert space is characterised by a reproducing kernel on $\H$, 
$K_p(x,y)$, which is a smooth, bounded, symmetric integral kernel that 
solves the equation $\nabla^2 K= -(p^2+1)K$ in both $x$ and $y$.  The 
formula for $K$ is given in Section \ref{eval}. 
The Lorentz group acts by translations on this Hilbert space, 
and the resulting representation is unitary and  
irreducible. 
  
The elementary operator $A\colon p_1 \otimes  
\cdots\otimes p_m \to q_1\otimes\cdots\otimes q_n$ is defined by  
\begin{equation}\label{vertexdefn}   f_1\otimes 
\cdots\otimes f_m\mapsto \frac1{2\pi^2}\int_H  
 K_{q_1}(y_1,z) \cdots K_{q_n}(y_n,z) f_1(z) \cdots f_m(z) \; \dd z 
\end{equation} 
The result of applying the operator $A$ is a function of the variables 
$y_1, \ldots, y_n$ in $H$.  However it is not clear that this function 
lies in the Hilbert space $q_1 \otimes\cdots\otimes q_n$.  For now, we  
simply consider the formal expressions that are obtained by composing  
these operators.  We postpone the question about whether these converge
to the consideration of closed diagrams (those with no free ends). 
  
The operator $A$ can be thought of as an integral operator, albeit with 
a distributional integral kernel, 
\begin{equation}\label{kerneldefn}  
 A(x_1,\ldots,x_m; y_1,\ldots, y_n) = \end{equation} 
\[  \frac1{2\pi^2}\int_H
K_{q_1}(y_1,z)\cdots K_{q_n}(y_n,z)
\delta(z,x_1)\cdots  \delta(z,x_m)  \; \dd z  \] 
with $\delta(x,y)$ the delta function for integration on $H$.  With this 
notation, the composition of operators consists of multiplying their 
kernels and integrating the common variables over $H$.  The trace is 
naturally expressed using the integral kernels in the same way.  For 
example, the trace of $A$ over the first variable is 
\[  \int_H A(z,x_2,\ldots,x_m; z,y_2,\ldots, y_n) \;\dd z. \] 
 
Combining the elementary operators as described above gives a 
description of the operator for a general diagram. Since composition and 
trace both involve joining a free end at the bottom of the diagram  
to a free end on the top, the calculation always gives a factor of 
\[ \int_H K_p(y_1,x) \delta(x,y_2)\; \dd x = K_p(y_1,y_2)  \] 
for each internal edge, with variables $y_1, y_2$ corresponding to  
its two vertices. 
 
The operator for a diagram with just two 
free ends $$\epsfysize=1.5in \epsfbox{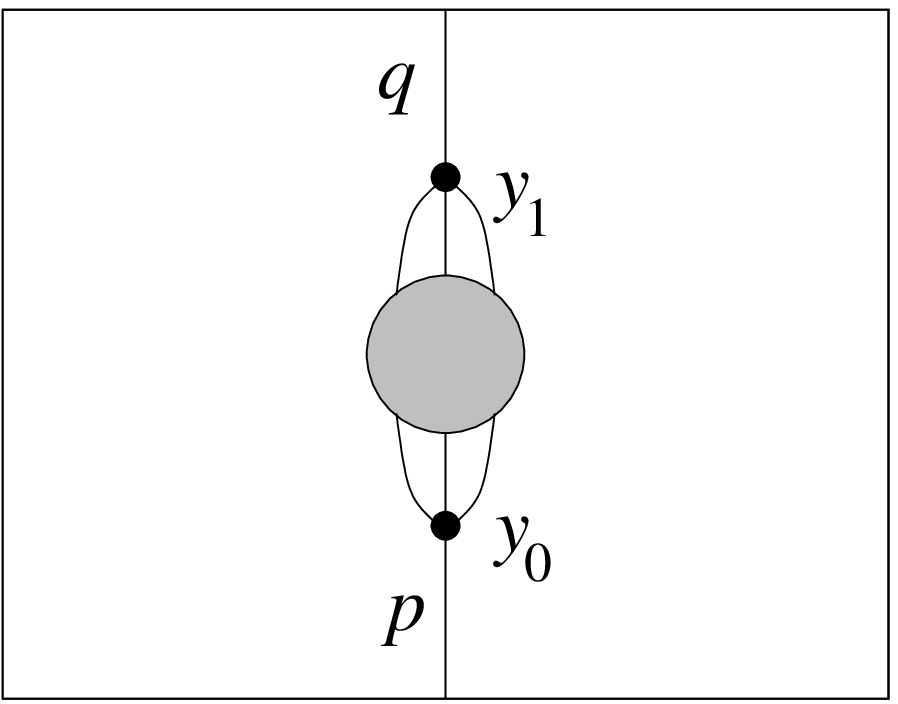}$$ can be described as 
follows. Associate a variable $y_i\in\H$ to each vertex in the 
diagram. Let $E$ be the set of all edges in the interior of the diagram 
(i.e., not meeting the boundary box). The corresponding operator is 
$$f\mapsto \int\prod_i\frac{\dd y_i}{2\pi^2} \left(\prod_E K\right) 
K_q(y_1,x)f(y_0).$$ 
  
In the main body of the paper we consider \textit{closed} diagrams, 
those with no free ends on the boundary rectangle. The naive idea for 
associating a number, or evaluation, to this graph would be to take 
the trace of the previous graph.  But this trace is always 
infinite. However, by Schur's lemma 
$$ \epsfysize=1.5in\epsfcenter{graph.eps}\quad = \quad \Lambda \quad \quad 
\epsfysize=1.5in\epsfcenter{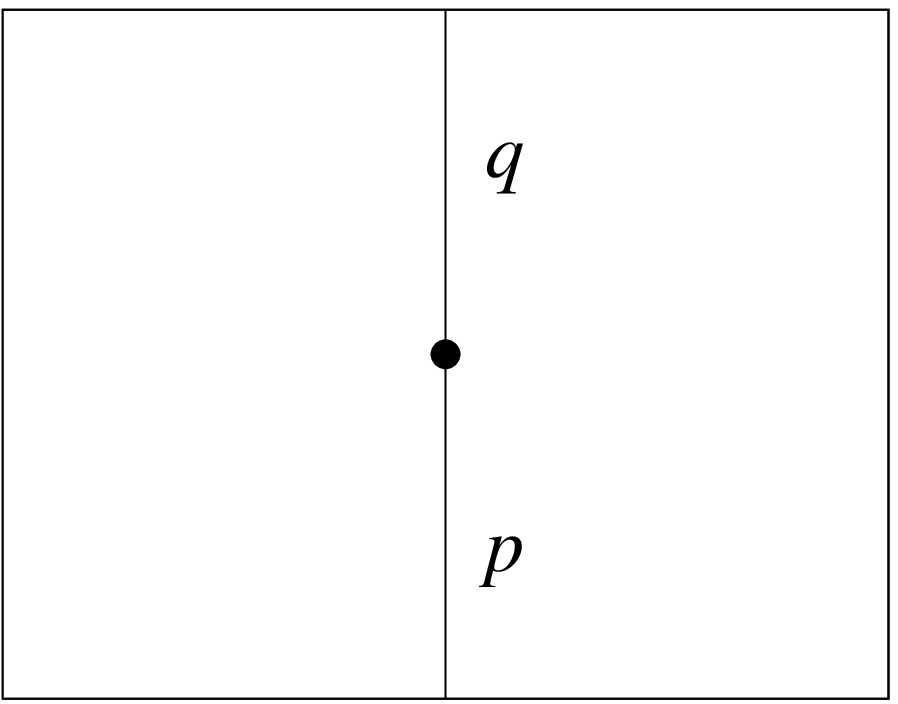}$$  
for some constant $\Lambda$,  
and in fact both sides are proportional to $\delta(p-q)$. 
 
In what follows, we show that a good definition for the evaluation 
of the closed diagram  
$$\epsfysize=1.5in\epsfbox{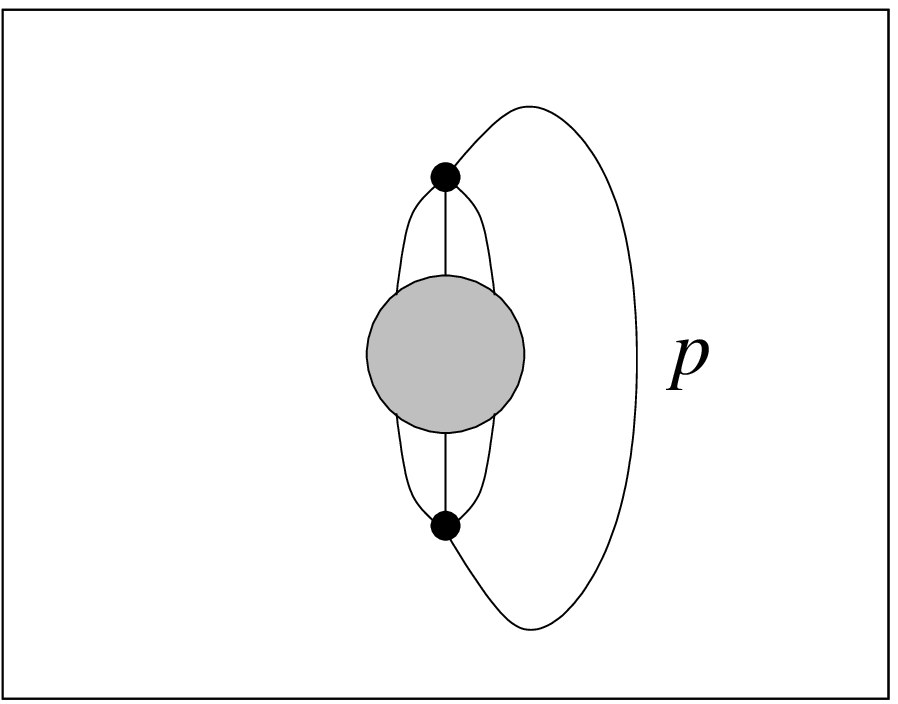}.$$ 
is the above constant $\Lambda$; this constant 
is given by a multiple integral of products of $K$'s which 
converges in many cases.  This definition is presented afresh in Section 
\ref{eval} with no reference to the representation theory sketched 
here. In that section, it is shown that this gives an invariant of the 
graph which does not depend on the way in which the diagram is drawn in 
the plane, nor on which edge is chosen to break it into a diagram with two 
free ends. 
  
\subsection{Physics applications} 
 
The study of spin networks was initiated by Penrose \cite{P} in the 
early 1970's, as part of an attempt to a find a description of the 
geometry of spacetime that takes quantum mechanics into account from the 
very start.  These spin networks were simply graphs with edges labelled 
by irreducible representations of $\SU(2)$ (i.e.\ spins $j = 0, {\frac 
1 2}, 1, {\frac 3 2}, \dots$) and vertices labelled by intertwining 
operators.  Such spin networks are also implicit in Ponzano and Regge's 
state sum model of 3-dimensional quantum gravity, published in 1968 
\cite{PonzReg}.  However, this was only realized much later \cite{HP}. 
 
The real surge of work on spin networks came in the early 1990's, when 
they were generalized to other groups and even quantum groups.  At this 
point, people began to use them systematically to construct topological 
quantum field theories.   For example, Reshetikhin and Turaev \cite{RT} 
used spin networks to give a purely combinatorial description of 
Chern-Simons theory and prove that it satisfies the Atiyah axioms for a 
3-dimensional TQFT.  Shortly thereafter, Turaev and Viro \cite{TV} used 
them to construct a $q$-deformed version of the Ponzano-Regge model and 
prove that it, too, is a TQFT.  We now recognize this theory as a 
Euclidean signature version of 3d quantum gravity with nonzero 
cosmological constant.  Later, Crane and Yetter \cite{CY} used spin 
network technology to construct a state sum model of a 4d TQFT.  This 
appears to be a quantization of $BF$ theory with cosmological constant 
term.   
 
One reason these topological quantum field theories are interesting is 
that they share some features with a physically more important but also 
far more problematic theory: 4-dimensional quantum gravity.  The success 
of spin network methods in constructing TQFTs prompted various attempts 
to apply spin networks to 4d quantum gravity. These attempts came from 
two main directions.  
 
The first was work on `loop quantum gravity', an approach to the 
canonical quantization of Einstein's equations.  Here   
Rovelli and Smolin \cite{RS} realized that $\SU(2)$ spin networks 
embedded in a 3-manifold representing space can serve as an explicit 
basis of kinematical states.  Using this basis they showed how to 
construct operators corresponding to observables such as the areas of 
surfaces and volumes of regions \cite{RS2}.  This work was soon made  
rigorous by Ashtekar, Lewandowski, Baez and others, and spin networks 
quickly became a standard tool in this field \cite{A,A2,A3,BZ,BZ2}.   
 
The second direction was work on state sum models of 4d quantum gravity. 
Models of this sort were proposed by Barrett and Crane, first in the 
Euclidean \cite{BC} and then in the Lorentzian signature \cite{BC2}.   
Their original models involve a triangulation of a fixed 4-manifold 
representing spacetime, but there is also great interest in more 
abstract `spin foam models', where there is no underlying manifold 
\cite{BZ3,BZ4,DFKP,PR}.  Again, these models come in both Euclidean and 
Lorentzian versions.  The Lorentzian versions are more physically 
realistic, but they involve extra difficulties due to the noncompactness 
of the Lorentz group $\SO_0(3,1)$, the solution of which forms the main topic 
of this paper. 
 
\section{Evaluating relativistic spin networks} \label{eval} 
 
As discussed informally in the introduction, a {\sl relativistic spin
network} is a graph with an assignment of a real number $p\ge 0$ to each
edge.  In what follows only relativistic spin networks whose underlying
graph is connected are considered.
 
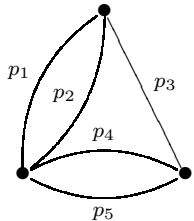
\begin{figure}[ht] 
$$\xymatrix{  
& *{\bullet}  
\ar@{-} 
@/^.75pc/ 
[ddl] 
_{p_2} 
\ar@{-} 
@/_.85pc/ 
[ddl] 
_{p_1}  
\ar@{-} 
[ddr] 
^{p_3}  
\\ \\ 
*{\bullet} 
\ar@{-} 
@/^.75pc/ 
[rr] 
^{p_4} 
\ar@{-} 
@/_.75pc/ 
[rr] 
_{p_5} 
&& 
*{\bullet} 
\\} 
$$ 
\caption{A relativistic spin network}\label{rsn}  
\end{figure} 
 
The idea of an evaluation is a function that gives a number for each 
relativistic spin network. In \cite{BC2} an integral was defined which 
determines a number for the relativistic spin network, if it 
converges.   The integral is based on the following kernel: 
\begin{equation}\label{zonal} 
K_p(x,y)=K_p(r)= \frac{\sin pr}{p\sinh r}  
\end{equation} 
where $x$ and $y$ are two points in three-dimensional hyperbolic space,
$\H$, and $r$ is the distance between them. This formula defines the
function $K_p(r)$ for $p,r > 0$, but it extends uniquely to a continuous
function of $p,r \ge 0$, with $K_p(0)=1$ and $K_0(r)=r/\sinh r$.
  
The integral is defined as follows.  First, to each vertex $v\in V$ of 
the graph we associate a variable $x_v \in \H$.  Each edge $e \in E$ 
thus has two variables, $x_{s(e)}$ and $x_{t(e)}$, associated to its 
endpoints.   Next, to each edge $e$ of the graph we associate a 
factor of $K_p(x_{s(e)},x_{t(e)})$, which depends on the edge label $p$.  
 
The idea is then to integrate the product of these factors $\prod_E K$ 
over the variables in hyperbolic space.  However, since this product is 
invariant under the action of $\SO_0(3,1)$ as isometries of $H$, one of the 
integrations is redundant, and would lead to an infinite value for the 
integral.   Thus we arbitrarily choose one vertex, say $w$, and omit the 
integration over the variable associated to that vertex.  Let $V' =  
V - \{w\}$ be the remaining set of vertices, and $n$ the total number of 
vertices. 
 
The integral is then given as follows: 
\begin{equation}\label{evaluation} 
I_w(x_w)= {\frac1 { ({2\pi^2})^{n-1}} } 
\int_{{\H}^{n-1}} \prod_E K_{p(e)}(x_{e(0)},x_{e(1)}) \prod_{v\in V'} \dd x_v 
\end{equation} 
The measure $\dd x$ on hyperbolic space is the standard Riemannian 
volume measure for the unit hyperboloid. In spherical coordinates where 
$r$ is the distance from a fixed origin  
\[ 
\dd x= \sinh^2r \,\dd r \,\dd \Omega, 
\] 
where $ \dd \Omega$ is Lebesgue measure on the unit 2-sphere. 
 
If the integral converges, it defines a function of the remaining 
variable, $x_w$.  However the Lorentz invariance gives $I_w(x_w)=I_w(L(x_w))$ 
for any $L \in \SO_0(3,1)$, so the integral is actually a constant, say 
$I_w$. 
 
The Lorentz invariance also implies that $I_w$ is independent of the 
choice of the vertex $w$. This follows from the formula 
\[ I_w(x_w)=\int_\H I(x_w,x_v) \,\dd x_v \] 
where the integrand is obtained by integrating over all but two 
variables. This function is an invariant function on $\H\times\H$, and 
therefore a function of the hyperbolic distance between its two 
arguments. Any such function is symmetric in its arguments. This 
establishes the equality $I_w(x_w)=I_v(x_v)$. 
 
\begin{defn} 
The {\rm evaluation} of a relativistic spin network is defined as 
the common value of $I_w(x_w)$ for any choice of vertex $w$ and $x_w \in \H$. 
\end{defn} 
 
Clearly this definition only works in the cases when the integral 
converges.  We have not yet been precise about what this means.  The best 
situation is when the integrand in the definition is integrable, i.e., 
when the Lebesgue integral of its absolute value exists.  In this case we 
say that the relativistic spin network is {\sl integrable}.  The 
results in this paper refer to the integrable case.  If the network is 
integrable for all values of the edge labels $p$, then the graph will be 
called integrable, and the evaluation defines a function of these 
edge labels.  Our first result is that this function is bounded: 
 
\begin{theorem}\label{bound} For an integrable graph, the relativistic 
spin network evaluation is bounded by a constant that is independent of 
the edge labels. \end{theorem} 
 
A more general situation is where the integral defines a generalised 
function, or distribution, in the $p$ variables.  In this situation the 
relativistic spin network may not be integrable for specific values of 
the edge labels $p$.  However, the integrand in the definition of the 
evaluation will become integrable after smoothing with suitable test 
functions in the $p$ variables.  We are not going to be precise about 
the details of the test functions, but will confine ourselves to noting 
some simple examples where this phenomenon occurs.  In this case, the 
graph will be called {\sl distributional}.  If the graph is not even 
distributional, it will be called {\sl divergent}. 
  
Some simple cases, calculated in \cite{BC2}, illustrate these 
definitions.  A graph given by a single loop with one vertex on it: 
$$\monogon p$$ 
has the evaluation 
\[ K_p(x,x)=1. \] 
 
A graph with two vertices on a loop: 
\[  \bigon ab \] 
has the evaluation 
\[ \nfactor \int_{\H} K_a(x,y) K_b(y,x)\,\dd y = 
\frac 2 {\pi ab} \int_0^\infty \sin ar \sin br\, \dd r 
=\frac{\delta(a-b)}{a^2}. \] 
This is an example of a relativistic spin network that is not integrable 
for any values of $a$ and $b$.  However the result is well defined after 
smoothing in $a$ and $b$, so the graph is distributional.  In general 
this phenomenon occurs whenever the graph has a bivalent vertex. 
  
In the simpler case of a closed network with two vertices joined with 
just one edge: 
\[  \unigon p \] 
the evaluation is a divergent integral for all values of the label $p$. 
\[ \nfactor \int_{\H} K_p(x,y)\,\dd y = \frac1 {2\pi^2 p} \int_0^\infty 
\sinh r\sin pr \,\dd r. \] 
More generally, any graph with a univalent vertex is divergent.  
 
The theta graph with two vertices and three edges: 
\[  \thetagraph abc \] 
is integrable and has the evaluation 
\begin{multline*}\nfactor \int_{\H} K_a(x,y) K_{b}(x,y) K_c(x,y)\,\dd y 
=\frac2{\pi abc}\int_0^\infty \frac{\sin ar\sin br\sin cr}{\sinh r}\,\dd r\\ 
=\frac1{4 abc} 
\bigl(\tanh(\frac\pi2(b+c-a))+ \tanh(\frac\pi2(c+a-b))\\ 
 + \tanh(\frac\pi2(a+b-c))- \tanh(\frac\pi2(a+b+c)) \bigr)\end{multline*} 
The graphs with two vertices and more than three connecting edges are 
also integrable and can be evaluated explicitly. 
 
Now we can extend these examples by stating our main results on 
integrability.   In what follows, we call the complete graph on  
4 vertices the `tetrahedron': 
$$\xymatrix{ *{\bullet} 
\ar@{-} 
[r]  
\ar@{-} 
[d] 
\ar@{-} 
[dr] 
& 
*{\bullet} 
\ar@{-} 
[d]  
\\ 
*{\bullet} 
\ar@{-} 
[r]  
\ar@{-} 
[ur]  
& 
*{\bullet} 
\\}$$ 
and call the complete graph on 5 vertices the `4-simplex': 
$$\xy/r3pc/: 
{\xypolygon5~*{\bullet}}, 
"1";"3"**@{-}, 
"2";"4"**@{-}, 
"3";"5"**@{-}, 
"4";"1"**@{-}, 
"5";"2"**@{-} 
\endxy 
$$ 
\medskip
 
\begin{theorem}\label{tetrahedron} The tetrahedron is an integrable 
graph.  \end{theorem} 
 
\begin{theorem}\label{construct} 
A graph obtained from an integrable graph by connecting an extra vertex  
to the existing graph by at least three extra edges is integrable.  A graph  
obtained from an integrable graph by adding extra edges is integrable.   
A graph constructed by joining two disjoint integrable graphs  
at a vertex is integrable. 
\end{theorem} 
 
\medskip 
\centerline{\epsfysize 1.2in \epsfbox{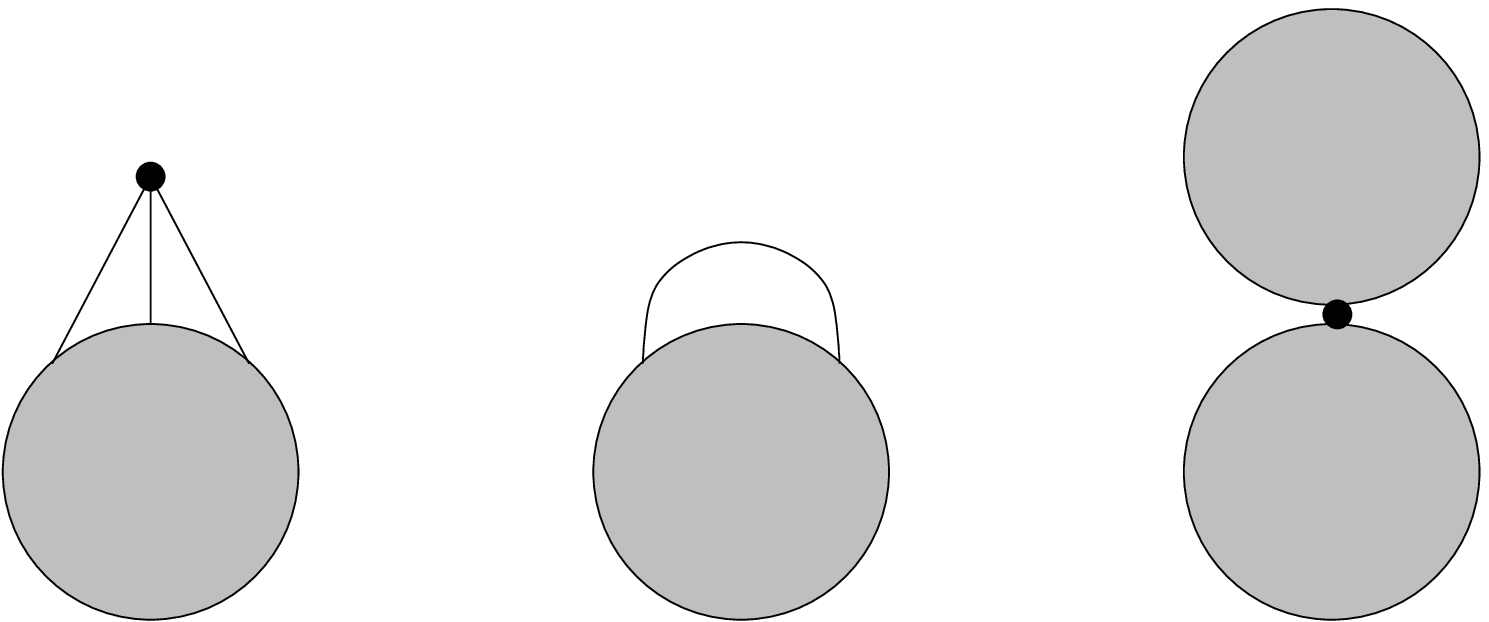}} 
\medskip 
 
Theorem \ref{construct} allows the construction of a large class of 
integrable graphs.  Starting with a graph which is known to be 
integrable, such as the theta graph or the tetrahedron, then one can 
construct larger  integrable graphs by successively carrying out the 
operations described in the theorem.  In particular, the 4-simplex is 
integrable because it is obtained by adding a vertex connected by four 
edges to the tetrahedron.    
 
\section{Proof of the Main Results}\label{proofs} 
 
In equation (\ref{evaluation}) we defined the evaluation of a 
relativistic spin network as a certain integral over $n$-tuples of 
points in hyperbolic space.  To prove Theorems \ref{tetrahedron} and  
\ref{construct}, we need to show this integral converges.  Before starting  
the proofs, we give an informal outline of the procedure. 
 
Our procedure is to integrate over one point (variable)      
at a time, treating the remaining ones as 
fixed.  This is justified by the theorems of Fubini and Tonelli concerning 
the Lebesgue integral of a function of several variables \cite{Royden}. 
 
For an example of this procedure, consider the integral 
\[   
\int_\H \dd x \, K_{p_1}(x,x_1) K_{p_2}(x,x_2) \cdots K_{p_n}(x,x_n) 
\] 
for fixed $x_1, \dots, x_n \in \H$.   To prove that this integral 
converges we need a bound on the kernel $K$ (proved below in Lemma 
\ref{kernel.bound}): for any $\epsilon > 0$, 
there exists a constant $c$ (independent of $p$) such that 
\[     |K_p(r)| \le c e^{-(1 - \epsilon)r} . \] 
Using this, it follows that the integral is bounded by 
\[ c^n \int_\H \dd x \,  e^{-(1 - \epsilon)(r_1 + \cdots + r_n)}  \] 
where $r_i = d(x,x_i)$.  Now suppose we can find a `barycentre' for the 
points $x_i$, that is, a point $b \in \H$ such that 
\[  r := d(x,b) \le \frac 1 n (r_1 + \cdots + r_n)  \] for all $x$. 
Then if we work in spherical coordinates centered at $b$, we see that 
the integral is bounded by 
\[ 4 \pi c^n \int_0^\infty e^{-(1-\epsilon)nr} \sinh^2 r \, \dd r. \] 
which converges for all $n \ge 3$ providing we pick $0<\epsilon<1/3$.   
 
This example illustrates the importance of adding 3 or more new edges 
for each new vertex in the graph.  To prove our main results, we now 
prove the above bound on the kernel $K$, construct the required 
barycentres, and give an improved version of the above estimate.  Finally, we  
give a careful treatement of the tetrahedron graph. 
\medskip 
 
We begin the formal proofs by bounding $K$. First note that  
$|{\sin pr}| \le pr$ so that for $p>0$ 
\begin{equation} 
\label{K} 
|K_p(r)| = \frac{|{\sin pr}|} {p \sinh r} \le \frac r {\sinh r} . 
\end{equation} 
This bound on $K_p(r)$ also holds when $r$ is zero, as long as we  
define $r/\sinh r$ to equal $1$ when $r = 0$. 
The right-hand side of the inequality is $K_0(r)$.  
 
\begin{proof}[Proof of Theorem \ref{bound}] 
By inequality \ref{K}, the evaluation of a relativistic spin network 
is bounded by the evaluation of the network with the edge label $p = 0$  
for each edge.  
\end{proof} 
  
Now we give the detailed estimates needed for Theorems \ref{tetrahedron}  
and \ref{construct}.  
\begin{lemma} \label{kernel.bound}  For any $\epsilon > 0$, there 
exists a constant $c$ such that 
\[     |K_p(r)| \le c e^{-(1 - \epsilon)r}  \] 
for all $p \ge 0$ and $r \ge 0$. Also, $|K_p(r)|\le 1$ for all $p,r$. 
\end{lemma} 
 
\begin{proof}    
Since the function $r/\sinh r$ is bounded  and  
is asymptotic to $2re^{-r}$ as $r \to +\infty$, 
for any $\epsilon > 0$ there exists $c$ with  
\[  \frac r {\sinh r} \le c e^{-(1 - \epsilon)r} . \]  
The second part follows because $r/\sinh r$ is bounded by 1. 
\end{proof} 
 
Next we construct a barycentre for any finite collection of points in 
hyperbolic space, beginning with the case of two points. 
 
\begin{lemma} \label{barycentre1} Suppose $p_1,p_2 \in \H$ and let $p$ 
be the midpoint of the geodesic from $p_1$ to $p_2$.  For any point $q 
\in \H$ we have  
\[           d(p,q) \le {\frac 1 2}(d(p_1,q) + d(p_2,q))  \] 
\end{lemma} 
 
\begin{proof} Using notation as in this picture: 
 
\medskip 
\centerline{\epsfysize=1.8in\epsfbox{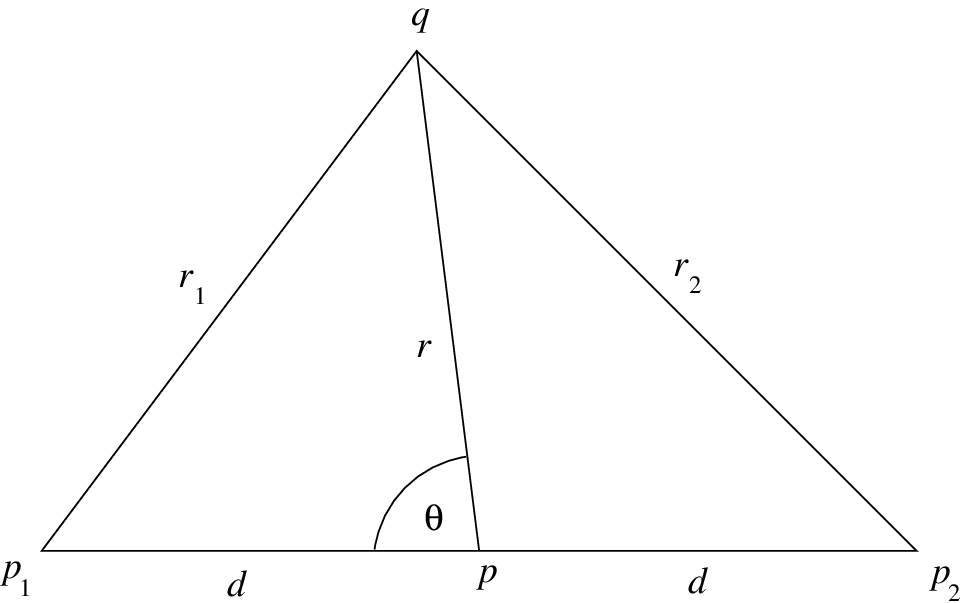}} 
\medskip  
 
\noindent we need to show $r \le \frac 1 2 (r_1 + r_2)$.   By the law of 
cosines for hyperbolic trigonometry \cite{Ratcliffe} we have 
\[    \cosh d \cosh r = \cosh r_1 + \cos \theta \sinh d \sinh r, \] 
\[    \cosh d \cosh r = \cosh r_2 - \cos \theta \sinh d \sinh r. \] 
Adding these equations we obtain 
\[  \cosh d \cosh r = \cosh({\frac{r_1 - r_2} 2})  
\cosh({\frac{r_1 + r_2}{2}}) .\] 
By the triangle inequality we have $ |r_1 - r_2| \le 2d$, so 
\[  \cosh r \le \cosh(\frac{r_1 + r_2} {2}), \] 
from which the desired result follows.    
\end{proof} 
  
There are a variety of ways of constructing a barycentre for 3 or more 
points.  First we give an intuitive method for 3 points, followed by an 
proof that a barycentre exists for any finite number of points. 
 
A barycentre for 3 points can be constructed by an iterative process. 
Begin by constructing midpoints of the geodesics between the points 
$p_1,p_2$ and  $p_3$ as in this picture: 
 
\medskip 
\centerline{\epsfysize=1.8in\epsfbox{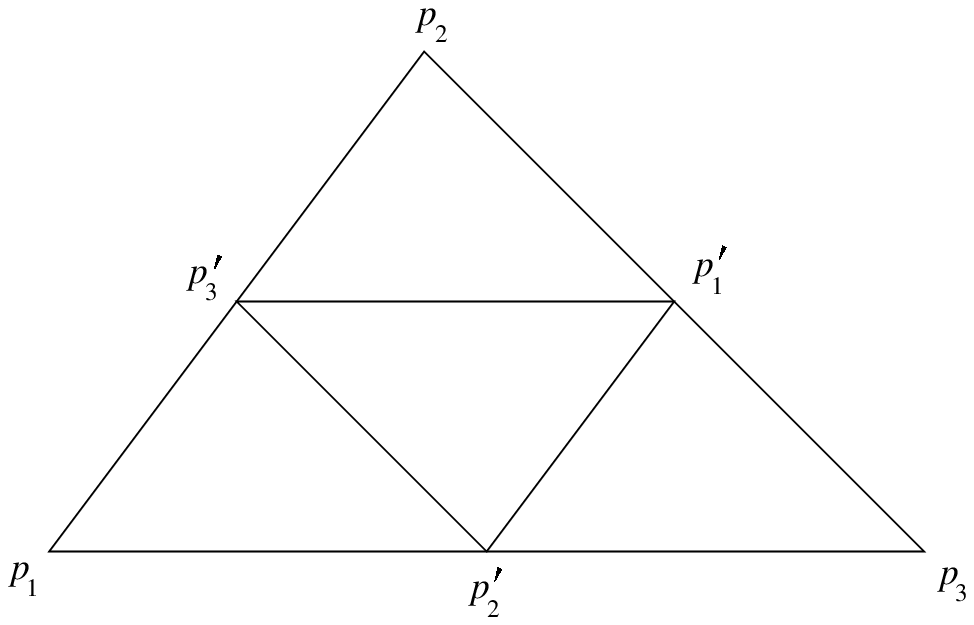}} 
\medskip 
 
\noindent Lemma \ref{barycentre1} implies the inequality  
\[      d(p_1,q) + d(p_2,q) + d(p_3,q) \ge  
        d(p'_1,q) + d(p'_2,q) + d(p'_3,q) .\] 
Iterating this process, we obtain a sequence of nested triangles 
in hyperbolic space: 
 
\medskip 
\centerline{\epsfysize=1.4in\epsfbox{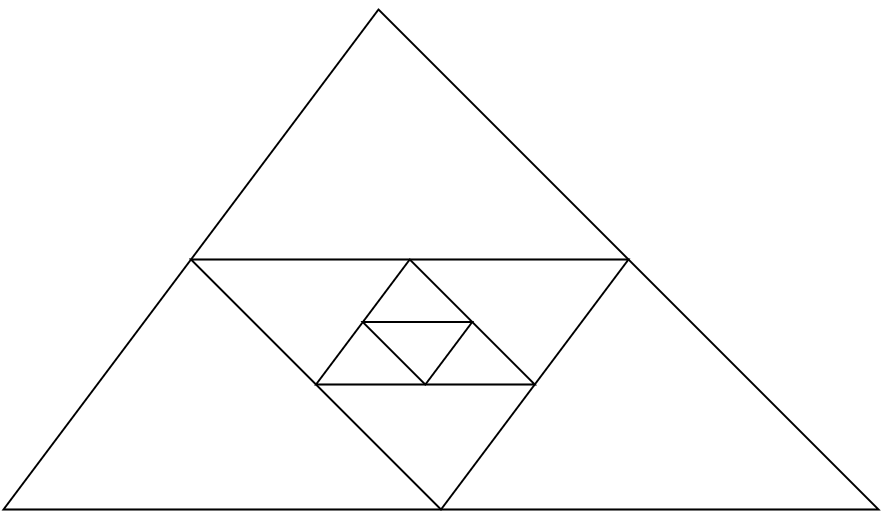}} 
\medskip 
 
\noindent The vertices converge to a point $p$, 
the unique point contained in all the triangles.   
By repeated use of the above inequality  
\[   3d(p,q) \le d(p_1,q) + d(p_2,q) + d(p_3,q) , \]           
so that $p$ is a barycentre. 
 
The following constructions work for all $n$. First, consider the case
where the points lie along a straight line, i.e.\ a geodesic
$\gamma\in\H$. Then since $\gamma$ is isometric to the real line with
its usual metric, the arithmetric mean of the points is defined. The
next lemma shows that this is a barycentre.
 
\begin{lemma} \label{barycentre3} Suppose $\gamma\subset \H$ is a
geodesic, and $p_1,p_2,\ldots,p_n \in \gamma$. Then the arithmetic mean
$p$ of the points has the property that for any point $q \in \H$ we have
\[ d(p,q) \le {\frac 1 n}(d(p_1,q) + d(p_2,q) + \cdots + d(p_n,q) ).  \]
\end{lemma}
 
\begin{proof} This is proved by iteration. Suppose $p_1$ and $p_2$ are
two points in the set which are the farthest distance apart. Then using
Lemma \ref{barycentre1}, we can replace both $p_1$ and $p_2$ by the
barycentre of $p_1$ and $p_2$ without increasing the quantity
\[          d(p_1,q) + d(p_2,q) + \cdots + d(p_n,q) .  \] 
By iterating this process, all of the points in the set converge to the
arithmetic mean $p$.  
\end{proof}
 
Now the main result on barycentres is proved. 
 
\begin{lemma} \label{barycentre2} Suppose $p_1,p_2,\ldots,p_n \in \H$. Then  
there exists a point $p$ such that for any point $q \in \H$ we have  
\[           d(p,q) \le {\frac 1 n}(d(p_1,q) + d(p_2,q) + \cdots + d(p_n,q) ).  \] 
\end{lemma} 
 
\begin{proof}   
This is proved by induction on $n$. The induction starts with $n=2$ by  
Lemma \ref{barycentre1}. Let $b$ be a barycentre for the first $n-1$ points.  
Then by the induction hypothesis 
\[    (n-1)    d(b,q) + d(p_n,q) \le d(p_1,q) + d(p_2,q) + \cdots + d(p_n,q).  
 \] 
Next we find a barycentre $p$ for $n-1$ points at $b$ and 1 point at  
$p_n$.  To do this, note that all these points lie on a geodesic, and so the barycentre is constructed by Lemma \ref{barycentre3}. 
We thus have 
\begin{eqnarray*}    
n d(p,q) &\le& (n-1)  d(b,q) + d(p_n,q) \\ 
&\le& d(p_1,q) + d(p_2,q) + \cdots + d(p_n,q).   
\end{eqnarray*} 
\end{proof} 
 
Next we prove an improved version of the estimate given at the  
beginning of this section. Suppose $x_1,\ldots,x_n$ are fixed in $H$  
and $r_{ij} = d(x_i,x_j)$. 
  
\begin{lemma} \label{inductive} 
If $n \ge 3$, the integral 
\[    
J = \int_\H \dd x \, |K_{p_1}(x,x_1) K_{p_2}(x,x_2) \cdots K_{p_n}(x,x_n)| 
\] 
converges, and for any $0<\epsilon < 1/3$ there exists a constant 
$C > 0$ such that for any choice of the points $x_1,\dots,x_n$,  
\[ 
J \le C\,  
\exp\left(-\frac{n-2-n\epsilon}{n(n-1)}\sum_{i<j}r_{ij}\right) 
\] 
\end{lemma} 
 
\begin{proof} Define $r_i = d(x_i,x)$. Using Lemma \ref{kernel.bound}  
we obtain 
\[   |K_{p_1}(x,x_1) \cdots K_{p_n}(x,x_n)|  
\le c^n e^{-(1-\epsilon)\sum r_i} . \] 
The summation here, and in the rest of the proof of the lemma,
is over values of $i$ from $1$ to $n$.     
If we work in spherical coordinates using the barycentre of the points 
$x_1,x_2,\ldots, x_n$ as our origin, this implies 
\[ J \le 4 \pi c^n \int_0^\infty e^{-(1-\epsilon)\sum r_i}  
\sinh^2 r \, \dd r .\] 
We next break this integral over $r$ into two parts, and estimate these 
separately using two bounds: 
\[       \sum r_i \ge nr  \] 
from Lemma \ref{barycentre2}, and  
\[           \sum r_i\ge nM  \] 
where  
\[      M = \frac 1 n\min_{x} \sum r_i. \] 
We obtain 
\begin{eqnarray*}  J &\le& 
4 \pi c^n 
\left[ \int_0^M e^{-n(1-\epsilon)M} \sinh^2 r \, \dd r +  
\int_M^\infty e^{-n(1-\epsilon)r} \sinh^2 r \, \dd r \right] \\ 
&\le&   \pi c^n           
\left[ \int_0^M e^{-n(1-\epsilon)M + 2r} \, \dd r +  
\int_M^\infty e^{-n(1 -\epsilon)r + 2r} \, \dd r \right] \\ 
&\le& C e^{-(n-2-n\epsilon)M}   
\end{eqnarray*} 
for some constant $C > 0$ depending only on $\epsilon$ and $n$.   Finally, the 
triangle inequality implies 
\[ \sum r_i \ge \frac 1 {n-1} \sum_{i<j}r_{ij} \] 
for all choices of $x_1,x_2,\ldots,x_n$, so  
\[  M \ge \frac 1 {n(n-1)}  \sum_{i<j}r_{ij}. \] 
\end{proof} 
 
\begin{proof}[Proof of Theorem \ref{construct}] 
Firstly, consider introducing an extra vertex connected to an integrable  
graph $\gamma$ by three or more extra edges. Lemma \ref{inductive} shows that  
doing the integral over the extra vertex first introduces an extra  
multiplicative factor of $J$ in the integral of $\prod_E |K|$ for the  
graph $\gamma$. However the function $J$ is bounded so the new graph is  
also integrable. 
  
Adding an extra edge does not affect the integrability of a graph  
because $|K|<1$, by Lemma \ref{kernel.bound}. 
    
Finally, for two integrable graphs joined by identifying one vertex we have  
that the evaluation of the resulting graph is the product of the evaluations  
of the two pieces, and so in particular the graph is integrable. This follows  
from taking the vertex where the two pieces are joined as the vertex which is  
not integrated over in the definition of the evaluation. 
 \end{proof} 
 
By using Lemma \ref{inductive} to evaluate a bound for the integral one
vertex at a time, one can actually prove that the $n$-simplex is
integrable for $n\ge5$. However to do the important cases of the
tetrahedron and the 4-simplex requires a more sensitive bound at the
stage where there are 3 vertices.
  
\begin{proof}[Proof of Theorem \ref{tetrahedron}] 
We need to show for any  
choice of numbers $p_{ij} \ge 0$ for $1 \le i < j \le 4$ and a  
point $x_1 \in \H$, the integral 
\begin{eqnarray*}   I &=& \int_{\H^3} \dd x_2 \,\dd x_3 \,\dd x_4 \; 
|K_{p_{12}}(x_1,x_2) K_{p_{13}}(x_1,x_3) K_{p_{14}}(x_1,x_4) \\  
&&  \qquad \qquad  
K_{p_{23}}(x_2,x_3) K_{p_{24}}(x_2,x_4) K_{p_{34}}(x_3,x_4)| 
\end{eqnarray*} 
converges.    
  
First we integrate out $x_4$ using Lemma \ref{inductive}, 
obtaining 
\[   I \le C \int_{\H^2} \dd x_2 \,\dd x_3 \;  
e^{-\frac 1 6 (1 - 3\epsilon)(r_{12} + r_{13} + r_{23})} \, 
|K_{p_{12}}(x_1,x_2) K_{p_{13}}(x_1,x_3) K_{p_{23}}(x_2,x_3)| \] 
where $r_{ij} = d(x_i,x_j)$.    
 
Next integrate over another variable, say 
$x_3$.  With 
\[  L = \int_\H \dd x_3 \; e^{-\frac 1 6 (1 - 3\epsilon)(r_{13} + r_{23})} \, 
|K_{p_{13}}(x_1,x_3) K_{p_{23}}(x_2,x_3)| \] 
this gives 
\begin{equation} 
\label{I} 
I \le  
C \int_\H \dd x_2 \; L \, e^{-\frac 1 6 (1 - 3\epsilon) r_{12}} \,  
|K_{p_{12}}(x_1,x_2)| . 
\end{equation}  
 
By equation (\ref{K}) we have 
\begin{eqnarray} 
L &\le& \int_\H \dd x_3 \; \frac{r_{13}r_{23}\, e^{-\frac 1 6 (1 - 3\epsilon) 
(r_{13} + r_{23})}}  
{\sinh r_{13} \sinh r_{23}} \nonumber \\ 
&\le& \int_\H \dd x_3 \;  
\frac{(r_{13} + r_{23})^2 e^{-\frac 1 6 (1 - 3\epsilon)(r_{13} + r_{23})} } 
{\sinh r_{13} \sinh r_{23}}  
\label{L} 
\end{eqnarray} 
 
To get a good bound on the integral here, we resort to a  
coordinate system in which two of the coordinates are  
\[ k  = \frac 1 2 (r_{13} + r_{23}), \qquad  
   \ell = \frac 1 2 (r_{13} - r_{23}), \] 
while the third is the angle $\phi$ between the plane 
containing $x_1,x_2,x_3$ and a given plane 
containing  $x_1$ and $x_2$.  The ranges of these 
coordinates are 
\[   r/2 \le k < \infty, \qquad   
     -r/2 \le \ell \le r/2 , \qquad     
     0 \le \phi < 2 \pi , \] 
where we set $r = r_{12}$.  Coordinates of this 
sort can also be defined in Euclidean space, where they are closely akin 
to prolate spheroidal coordinates \cite{MoonSpencer}, but here all 
the formulas are a bit different, since we are working in hyperbolic 
space.  The main thing we need is a formula for the volume form 
in these coordinates, 
\[ \dd x_3 =  
2\,\frac{\sinh r_{13} \, \sinh r_{23}} {\sinh r} \; \dd k \, \dd \ell \,  
\dd \phi \]    
which is proved in the Appendix. 
 
Using this formula we can do the integral 
(\ref{L}) in the $(k,\ell,\phi)$ coordinate system, obtaining 
\[ L \le 8 \int_0^{2 \pi} \dd \phi \int_{r/2}^\infty  
\dd k \int_{-r/2}^{r/2} \dd \ell \;  
\frac{k^2 e^{-\frac 1 3 (1 - 3\epsilon) k}} {\sinh r} \] 
or doing the integral over $\phi$ and $\ell$ and then $k$,  
\begin{eqnarray} 
 L &\le&  \frac{16 \pi r} {\sinh r}  
\int_{r/2}^\infty k^2 e^{-\frac 1 3 (1 - 3\epsilon) k} \, \dd k \nonumber \\ 
&\le&  \frac{(Ar^3 + B) e^{-{\frac 1 6 (1 - 3\epsilon) r}}} {\sinh r}  
\label{L2} 
\end{eqnarray} 
for some constants $A$ and $B$ independent of all the parameters 
in this problem. 
 
We conclude the proof by using this bound on $L$ to bound the 
integral $I$.   By (\ref{K}) and (\ref{I}) we have 
\begin{eqnarray*} 
I &\le& 
C \int_\H \dd x_2 \; L \, e^{- \frac 1 6 (1 - 3\epsilon) r} 
\, |K_{p_{12}}(x_1,x_2)|  \\ 
&\le& 4 \pi C \int_0^\infty L\, r e^{-\frac 1 6 (1 - 3\epsilon) r} \sinh r  
\, \dd r  
\end{eqnarray*} 
and by (\ref{L2}) this gives 
\[ I \le 4 \pi C  
\int_0^\infty   r (Ar^3 + B) e^{-\frac 1 3 (1 - 3 \epsilon) r} \, \dd r.  \] 
The right-hand side is finite so the proof is complete.   
\end{proof} 
  
\section{Remarks and Conclusions} 
 
Theorem \ref{construct} gives a large class of integrable graphs,
starting with the theta and tetrahedron graphs. However there are
further examples of integrable graphs. For example, the graph in Figure
\ref{rsn} is also integrable, but cannot be constructed from any
integrable graph by the methods of Theorem \ref{construct}. Its
integrability follows by applying Lemma \ref{inductive} to one of the
trivalent vertices.
 
It seems reasonable to conjecture that any 3-edge-connected graph is
integrable.  A 3-edge-connected graph is one that remains connected when
any edge is removed or any pair of edges are removed.  
 
It seems that the bound in Theorem \ref{bound} should be dramatically 
improved. Indeed, $K$ satisfies the bound $|K|<1/(p\sinh r)$ for 
$r>1/p$, thus for large $p$ one expects the evaluation to behave like 
$1/p$ for each edge  variable. By inspection, this is the case for the 
theta-graph.  We conjecture that a similar bound is true for all graphs 
not containing edge-loops (edges with both ends at the same vertex). 
 
It would also be interesting to consider the obvious generalization 
of this theory to other dimensions.  For applications to quantum
gravity, one would want the $(n+1)$-simplex to be an integrable graph
when labelled by any representations of $\SO_0(n,1)$ appearing in the
direct integral decomposition of the space of $L^2$ functions on 
$n$-dimensional hyperbolic space.

\section{Appendix: Spheroidal Coordinates in Hyperbolic Space} 
 
If we fix two points $x_1,x_2$ in three-dimensional hyperbolic 
space, and pick a hyperbolic plane containing the geodesic between 
these points, we can define {\sl spheroidal coordinates} on  
hyperbolic space as follows.  Given any point $x_3$ in hyperbolic 
space, its first two coordinates are 
\[ k  = \frac 1 2 (r_{13} + r_{23}), \qquad  
   \ell = \frac 1 2 (r_{13} - r_{23}), \] 
where $r_{ij}$ is the distance from $x_i$ to $x_j$.  The third 
coordinate is the angle $\phi$ between the plane containing 
$x_1, x_2, x_3$ and the given plane 
containing $x_1$ and $x_2$.  The ranges of  
these coordinates are 
\[   r/2 \le k < \infty, \qquad   
     -r/2 \le \ell \le r/2 , \qquad     
     0 \le \phi < 2 \pi , \] 
where we set $r = r_{12}$.   
 
In these coordinates, the volume form on hyperbolic 
space is given by 
\[ \dd x_3 =  
2\,\frac{\sinh r_{13} \, \sinh r_{23}} {\sinh r} \; \dd k \, \dd \ell \,  
\dd \phi. \]    
To prove this, it is easiest to consult the following 
picture and use the method of infinitesimals (or differential forms): 
 
\medskip 
\centerline{\epsfysize=2.0in\epsfbox{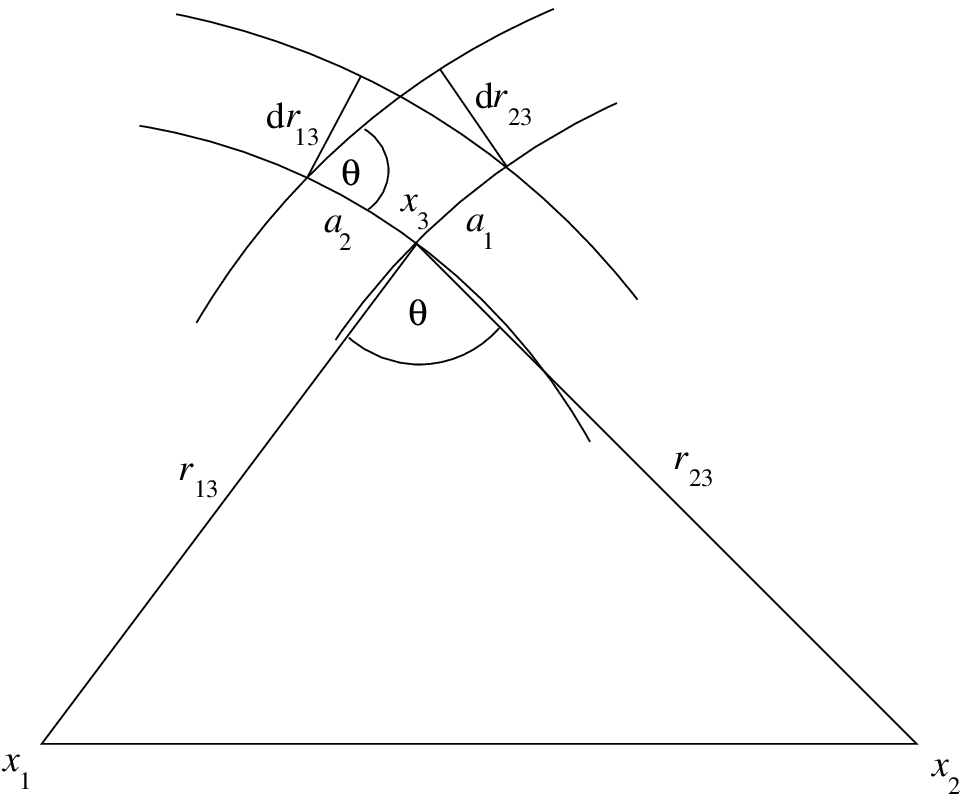}} 
\medskip 
 
\noindent The area of the infinitesimal parallelogram formed as we 
vary $r_{13}$ and $r_{23}$ by amounts $\dd r_{13}$ and $\dd r_{23}$ is  
$\sin \theta  \, a_1 a_2$, where $\theta$ is the angle  
between the geodesics from $x_3$ to $x_1$ and $x_2$.   
Evidently $\sin \theta \, a_i = \dd r_{i3}$, so this  
area is $\dd r_{13} \, \dd r_{23}/\sin \theta = 2 \dd k \,  
\dd \ell/\sin \theta$.   
As we vary $\phi$ by an amount $\dd \phi$,  
this parallelogram sweeps
out an infinitesimal paralleliped of volume  
\[ \dd x_3 = 2\, \frac {\sinh y} {\sin \theta} \, \dd k \, \dd \ell  \, 
\dd \phi , \] 
where $y$ is the distance from $x_3$ to the geodesic between $x_1$ 
and $x_2$.  With the help of the following picture: 
 
\medskip 
\centerline{\epsfysize=1.8in\epsfbox{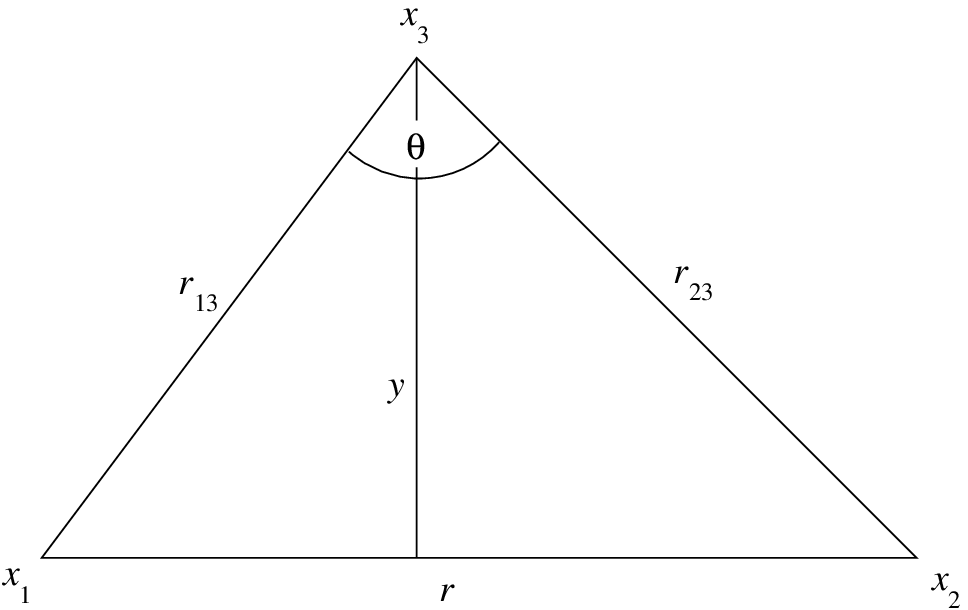}} 
\medskip 
 
\noindent repeated use of the hyperbolic law of sines gives 
\[  \sinh y =  
\frac{\sinh r_{13} \, \sinh r_{23}} {\sinh r} \; \sin\theta \] 
and thus 
\[ \dd x_3 =  
2\,\frac{\sinh r_{13} \, \sinh r_{23}} {\sinh r} \; \dd k \, \dd \ell \,  
\dd \phi. \]

\end{document}